\documentclass[12pt]{article}

\usepackage{amsmath}
\usepackage{subfigure}
\usepackage{appendix,epic,eepic,amsmath,amsthm,amssymb,epsfig,cite,indentfirst,oldgerm}
\renewcommand{\theequation}{\arabic{equation}}
\newcommand{\EQ}{\begin{equation}}
\newcommand{\EN}{\end{equation}}

\newcommand{\ket}[1]{\left|#1\right\rangle}      % Ket-Zustand
\newcommand{\IM}{\mathbf{\imath}}

\newcommand{\bear}{\begin{eqnarray}}
\newcommand{\ear}{\end{eqnarray}}
\newcommand{\bt} { \begin{tabular} }
\newcommand{\et}{ \end{tabular} }
\newcommand{\bc} { \begin{center} }
\newcommand{\ec}{ \end{center} }

\newcommand{\btb} { \begin{table} }
\newcommand{\etb}{ \end{table} }
\begin{document}

\topmargin 0pt
\oddsidemargin 5mm
\newcommand{\NP}[1]{Nucl.\ Phys.\ {\bf #1}}
\newcommand{\PL}[1]{Phys.\ Lett.\ {\bf #1}}
\newcommand{\NC}[1]{Nuovo Cimento {\bf #1}}
\newcommand{\CMP}[1]{Comm.\ Math.\ Phys.\ {\bf #1}}
\newcommand{\PR}[1]{Phys.\ Rev.\ {\bf #1}}
\newcommand{\PRL}[1]{Phys.\ Rev.\ Lett.\ {\bf #1}}
\newcommand{\MPL}[1]{Mod.\ Phys.\ Lett.\ {\bf #1}}
\newcommand{\JETP}[1]{Sov.\ Phys.\ JETP {\bf #1}}
\newcommand{\TMP}[1]{Teor.\ Mat.\ Fiz.\ {\bf #1}}

\renewcommand{\thefootnote}{\fnsymbol{footnote}}

\newpage
\setcounter{page}{0}
\begin{titlepage}
\begin{flushright}

\end{flushright}
\vspace{0.5cm}
\begin{center}
{\large The spectrum of a vertex model and related spin one chain sitting
in a genus five curve.}\\
\vspace{1cm}
{\large M.J. Martins } \\
\vspace{0.15cm}
{\em Universidade Federal de S\~ao Carlos\\
Departamento de F\'{\i}sica \\
C.P. 676, 13565-905, S\~ao Carlos (SP), Brazil}\\
%E-mail Address: {\tt martins@df.ufscar.br}}\\
\vspace{0.35cm}
\end{center}
\vspace{0.5cm}

\begin{abstract}
We derive the transfer matrix eigenvalues of a three-state vertex model whose 
weights are based on
a $\mathrm{R}$-matrix not of difference form with 
spectral parameters
lying on a genus five curve. We have shown that the basic building blocks
for both the transfer matrix eigenvalues and Bethe equations can
be expressed in terms of meromorphic functions 
on an elliptic curve. 
We discuss the properties of an underlying spin one chain originated
from a particular choice of the $\mathrm{R}$-matrix 
second spectral parameter. We present
numerical and analytical evidences that the respective low-energy
excitations can be gapped or massless depending on the strength
of the interaction coupling. In the massive phase we provide analytical 
and numerical evidences in favor of
an exact expression for the lowest energy gap.
We point out that 
the critical point separating these
two distinct physical regimes coincides with the 
one in which the weights geometry degenerate into union of genus one curves. 
\end{abstract}

\vspace{.15cm} \centerline{}
\vspace{.1cm} \centerline{Keywords: spin one chain, vertex model,
high genus curve.}
\vspace{.15cm} \centerline{June 2017}

\end{titlepage}

%\tableofcontents

\pagestyle{empty}

\newpage

\pagestyle{plain}
\pagenumbering{arabic}

\renewcommand{\thefootnote}{\arabic{footnote}}
\newtheorem{proposition}{Proposition}
\newtheorem{pr}{Proposition}
\newtheorem{remark}{Remark}
\newtheorem{re}{Remark}
\newtheorem{theorem}{Theorem}
\newtheorem{theo}{Theorem}

\def\ll{\left\lgroup}
\def\rr{\right\rgroup}

\newtheorem{Theorem}{Theorem}[section]
\newtheorem{Corollary}[Theorem]{Corollary}
\newtheorem{Proposition}[Theorem]{Proposition}
\newtheorem{Conjecture}[Theorem]{Conjecture}
\newtheorem{Lemma}[Theorem]{Lemma}
\newtheorem{Example}[Theorem]{Example}
\newtheorem{Note}[Theorem]{Note}
\newtheorem{Definition}[Theorem]{Definition}

\section{Introduction}

The $\mathrm{R}$-matrix is one of the  
basic elements in the theory of classical
two-dimensional integrable vertex model 
of statistical mechanics. In general, it is
a numerical
matrix of size $N^2 \times N^2$ depending 
on two independent spectral 
parameters $\lambda$ and $\mu$ denoted
here by $\mathrm{R}(\lambda,\mu) \in 
\mathbb{C}^N \otimes \mathbb{C}^{N}$. 
The integer $N$ 
is related to the number of possible statistical 
configurations at the edges of the vertex model. For integrability
it is sufficient to impose that the $\mathrm{R}$-matrix
fulfill the Yang-Baxter equation \cite{BAX},
\begin{equation}
\label{YB}
\mathrm{R}_{12}(\lambda_1,\lambda_2)
\mathrm{R}_{13}(\lambda_1,\lambda_3)
\mathrm{R}_{23}(\lambda_2,\lambda_3)=
\mathrm{R}_{23}(\lambda_2,\lambda_3)
\mathrm{R}_{13}(\lambda_1,\lambda_3)
\mathrm{R}_{12}(\lambda_1,\lambda_2),
\end{equation}
where $\mathrm{R}_{ab}(\lambda_a,\lambda_b) \in \mathbb{C}^{N} \otimes \mathbb{C}^{N} \otimes \mathbb{C}^{N}$ 
acts as the matrix $\mathrm{R}(\lambda_a,\lambda_b)$
on the $a$-th and $b$-th subspaces and as the identity 
on the remaining subspace component.

From a given $\mathrm{R}$-matrix one can always 
build an integrable 
vertex model interpreting
the matrices elements as the Boltzmann weights attached 
to the four edges of a vertex. The respective 
row-to-row transfer matrix
on a square lattice of size $\mathrm{L} \times \mathrm{L}$ 
can formally be written as an ordered product 
of $\mathrm{R}$-matrices \cite{ESTAFA},namely
\begin{equation}
\label{TRA}
\mathrm{T}(\lambda)=\mathrm{Tr}_0\left[ 
\mathrm{R}_{01}(\lambda,\mu)
\mathrm{R}_{02}(\lambda,\mu) \dots 
\mathrm{R}_{0\mathrm{L}}(\lambda,\mu)\right], 
\end{equation}
where the symbol $0$ denotes the auxiliary 
space associated to the horizontal degrees of freedom. The trace is taken on
such space and 
the second spectral parameter $\mu$ plays 
the role of an additional coupling of the model.

This vertex model generates a family of local Hamiltonians once
we consider that the $\mathrm{R}$-matrix is regular at some point
$\lambda_0$. This is to say that $\mathrm{R}(\lambda_0,\mu)$ is
proportional to the permutator 
$\mathrm{P}=\sum_{a,b=1}^N e_{ab} \otimes e_{ba}$ where $e_{ab}$ are the
$N \times N$ Weyl matrices. It follows that the logarithmic 
expansion of the transfer matrix (\ref{TRA}) around the point $\lambda_0$
generates a sequence of local commuting operators \cite{LUS,TATAFA} and the
Hamiltonian is the first non-trivial charge,
\begin{equation}
\label{HAM}
\mathrm{H} =\sum_{j=1}^{\mathrm{L}} \mathrm{P}_{j,j+1} \frac{\partial}{\partial \lambda} \mathrm{R}_{j,j+1}(\lambda,\mu)|_{\lambda=\lambda_0},
\end{equation}
where periodic boundary condition is assumed.

The physical understanding of vertex models needs the 
diagonalization
of their transfer matrices which are able to provide 
us the free energy 
and the nature of the low-energy excitations. 
The majority of the results on that matter has been 
concentrated in the particular situation
of $\mathrm{R}$-matrices depending on the difference of
the spectral variables.  
The exact diagonalization of the transfer matrix of vertex 
models whose $\mathrm{R}$-matrices are not of difference 
form are in fact more rare. As a notable exception one could mention 
is the covering vertex model associated to the one-dimensional
Hubbard Hamiltonian \cite{SHA,MAPB}.
The purpose of this
paper is to expand the number of examples in the literature 
dealing with the spectrum
of transfer matrices based on non-additive $\mathrm{R}$-matrices.
To this end we will investigate the structure of 
the transfer matrix eigenvalues
of a three-state vertex model with weights
lying on a genus five curve proposed recently 
by the author \cite{MAR}. More specifically, 
we shall consider the Yang-Baxter solution 
denominated in \cite{MAR}
special branch case whose underlying
curve $\mathrm{C}$ has the following affine form,
\begin{equation}
\label{CURVE}
\mathrm{C}=(x^2+\frac{y^2}{\varepsilon})(x^2+\varepsilon y^2)^2+\mathrm{U}\sqrt{\varepsilon}xy(x^2+\varepsilon y^2)-x^2+y^2=0,
\end{equation}
where $x$ and $y$ are the curve affine variables, 
the parameter $\mathrm{U}$ is free while $\varepsilon=\exp(\pm \IM \pi/3)$.
The basic structure of the $\mathrm{R}$-matrix up to 
an overall normalization is given by,
\EQ
\scriptsize{
\mathrm{R}(\lambda_1,\lambda_2)=\left[
\begin{array}{ccc|ccc|ccc}
 {a}(\lambda_1,\lambda_2) & 0 & 0 & 0 & 0 & 0 & 0 & 0 & 0 \\
 0 & {b}(\lambda_1,\lambda_2) & 0 & 1 & 0 & 0 & 0 & 0 & 0 \\
 0 & 0 & {f} (\lambda_1,\lambda_2) & 0 & {d}(\lambda_1,\lambda_2) & 0 & {h}(\lambda_1,\lambda_2) & 0 & 0 \\ \hline
 0 & 1 & 0 & \overline{b}(\lambda_1,\lambda_2) & 0 & 0 & 0 & 0 & 0 \\
 0 & 0 & \varepsilon {d} (\lambda_1,\lambda_2) & 0 & {g}(\lambda_1,\lambda_2) & 0 & {d}(\lambda_1,\lambda_2) & 0 & 0 \\
 0 & 0 & 0 & 0 & 0 & \overline{b}(\lambda_1,\lambda_2) & 0 & 1 & 0 \\\hline
 0 & 0 & \overline{h}(\lambda_1,\lambda_2) & 0 & \varepsilon {d}(\lambda_1,\lambda_2) & 0 & {f}(\lambda_1,\lambda_2) & 0 & 0 \\
 0 & 0 & 0 & 0 & 0 & 1 & 0 & {b}(\lambda_1,\lambda_2) & 0 \\
 0 & 0 & 0 & 0 & 0 & 0 & 0 & 0 & {a}(\lambda_1,\lambda_2) \\
\end{array}
\right].}
\label{RMA}
\EN

For the purposes of this paper we have performed  
a number of simplifications on the expressions of 
the $\mathrm{R}$-matrix elements
originally presented in \cite{MAR}. It turns out that they 
can be rewritten as follows,
\begin{eqnarray}
\label{WEI}
&& a(\lambda_1,\lambda_2)=\frac{x_1x_2}{x_2^2+\varepsilon y_2^2}+\varepsilon \frac{y_1y_2}{x_1^2+\varepsilon y_1^2},~~
b(\lambda_1,\lambda_2)=\frac{y_1x_2}{x_2^2+\varepsilon y_2^2}-\frac{x_1y_2}{x_1^2+\varepsilon y_1^2},~~
\overline{b}(\lambda_1,\lambda_2)=\frac{\varepsilon y_1x_2}{x_1^2+\varepsilon y_1^2}-\frac{\varepsilon x_1y_2}{x_2^2+\varepsilon y_2^2}
,\nonumber \\ \nonumber \\
&& d(\lambda_1,\lambda_2)=\frac{x_1y_1\left[x_2^2-y_2^2\right]}{\left[x_2^2+\varepsilon y_2^2\right]\left[x_1^2x_2^2-\varepsilon^2y_1^2y_2^2\right]}-\frac{x_2y_2\left[x_1^2-y_1^2\right]}{\left[x_1^2+\varepsilon y_1^2\right]\left[x_1^2x_2^2-\varepsilon^2 y_1^2y_2^2\right]}, \nonumber \\ \nonumber \\
&& f(\lambda_1,\lambda_2)=\frac{\varepsilon^2 x_1x_2y_2^2\left[x_1^2+\varepsilon y_1^2\right]}{\left[x_2^2+\varepsilon y_2^2\right]\left[x_1^2 x_2^2-\varepsilon^2 y_1^2 y_2^2\right]}
-\frac{x_1^2y_1y_2\left[x_2^2+\varepsilon y_2^2\right]}{\left[x_1^2+\varepsilon y_1^2\right]\left[x_1^2 x_2^2-\varepsilon^2 y_1^2 y_2^2\right]}
+\frac{y_1x_2\left[x_1y_1-\varepsilon^2 x_2y_2\right]}{x_1^2x_2^2-\varepsilon^2 y_1^2y_2^2}, \nonumber \\ \nonumber \\
&&g(\lambda_1,\lambda_2)=\frac{1+\varepsilon \left[d(\lambda_1,\lambda_2)\right]^2
-b(\lambda_1,\lambda_2)\overline{b}(\lambda_1,\lambda_2)}{a(\lambda_1,\lambda_2)+f(\lambda_1,\lambda_2)}, \nonumber \\ \nonumber \\
&&h(\lambda_1,\lambda_2)=a(\lambda_1,\lambda_2)+f(\lambda_1,\lambda_2)/\varepsilon,~~
\overline{h}(\lambda_1,\lambda_2)=a(\lambda_1,\lambda_2)+\varepsilon f(\lambda_1,\lambda_2),
\end{eqnarray}
where the subscript index $j$ 
denotes the point $x_j$ and $y_j$ on the genus five curve (\ref{CURVE}). 

In next section we discuss the diagonalization of the
transfer matrix (\ref{TRA},\ref{RMA}) within 
the algebraic Bethe framework. We have been able 
to show the part of the eigenvalues structure can be rewritten 
in terms of meromorphic functions depending
on auxiliary variables constrained by an 
elliptic curve. In section \ref{sec3} we study
the low-energy properties of a quantum spin one chain associated to a special 
choice of the second $\mathrm{R}$-matrix spectral
parameter. We investigate the spectrum by using the Bethe ansatz solution as well as 
the exact diagonalization of the Hamiltonian for small lattice sizes.
We present analytical and numerical evidences
that such spin chain
has massive excitations in the 
parameter $|\mathrm{U}| > 2 \sqrt{3}$ range and propose an expression for the lowest
mass gap.
In the regime $-2\sqrt{3} \leq \mathrm{U} \leq 2\sqrt{3}$ 
the results obtained from the exact diagonalization  support the view
that the low-energy behaviour should be massless. Our concluding remarks are summarized 
in section \ref{sec4}.

\section{The transfer matrix eigenvalues}

We can diagonalize the transfer matrix (\ref{TRA},\ref{RMA}) by means of the 
of algebraic Bethe ansatz method \cite{KO}. This is the case since
the $\mathrm{R}$-matrix commutes with the azimuthal component of an operator
with spin one. We can then choose the standard ferromagnetic
vacuum as the reference state in order to built the other eigenstates in sectors
where the total azimuthal magnetization is an arbitrary integer $n$.
We recall that this construction has been already performed 
in the work \cite{CAMA}
for the rather generic case of $\mathrm{R}$-matrices that are 
not of difference form. For a summary of the technical details entering
the construction of the
eigenvectors we refer to Appendix A. In what follows we present the final results
for the transfer matrix eigenvalues and denoting  them
by $\Lambda(\lambda,\mu)$ we obtain,
\begin{equation}
\Lambda(\lambda,\mu)= \left[a(\lambda,\mu)\right]^{\mathrm{L}} \prod_{i=1}^{\mathrm{L}-n}
\frac{a(\lambda_i,\lambda)}{\overline{b}(\lambda_i,\lambda)}
+\left[{\overline{b}}(\lambda,\mu)\right]^{\mathrm{L}} \prod_{i=1}^{\mathrm{L}-n}
\frac{\theta(\lambda,\lambda_i)a(\lambda,\lambda_i)}{{\overline{b}}(\lambda,\lambda_i)}
+\left[f(\lambda,\mu)\right]^{\mathrm{L}} \prod_{i=1}^{\mathrm{L}-n}
\frac{b(\lambda,\lambda_i)}{f(\lambda,\lambda_i)},
\end{equation}
where the phase shift $\theta(\lambda,\lambda_i)$ is defined as,
\begin{equation}
\label{teta}
\theta(\lambda,\lambda_i)=\frac{g(\lambda,\lambda_i)f(\lambda,\lambda_i)-\varepsilon \left[d(\lambda,\lambda_i)\right]^2}{a(\lambda,\lambda_i)f(\lambda,\lambda_i)}.
\end{equation}

As usual the rapidities $\lambda_j$ have to be chosen to satisfy certain non-linear constraints
named Bethe equations. In our case they are,
\begin{equation}
\label{BET}
\left[\frac{a(\lambda_j,\mu)}{\overline{b}(\lambda_j,\mu)}\right]^{\mathrm{L}}= 
\prod_{\stackrel{i=1}{i \neq j}}^{\mathrm{L}-n}
\frac{\theta(\lambda_j,\lambda_i)a(\lambda_j,\lambda_i)\overline{b}(\lambda_i,\lambda_j)}{a(\lambda_i,\lambda_j)\overline{b}(\lambda_j,\lambda_i)},~~j=1,\dots,\mathrm{L}-n.
\end{equation}

Direct inspection of the above formulae tells us that both 
the transfer matrix eigenvalues and the Bethe
ansatz equations depend on specific ratios of the weights. These ratios represent only a small subset  of
the field of fractions of a curve with genus five and they may have alternative representations on lower genus
curves. In what follows we investigate 
whether or not such fractions can 
be expressed in a simpler way with the help of new auxiliary
variables lying on other algebraic curve. In this sense, we first note that the simplest such ratio can be rewritten as,
\begin{equation}
\frac{a(\lambda_i,\lambda)}{\overline{b}(\lambda_i,\lambda)}= \left( \frac{x}{\varepsilon y} \right)
\left( \frac{\frac{x_i\left[x_i^2+\varepsilon y_i^2\right]}{y_i} +\varepsilon\frac{y\left[x^2+\varepsilon y^2\right]}{x}}
{\frac{x\left[x^2+\varepsilon y\right]}{y} -\frac{x_i\left[x_i^2+\varepsilon y_i^2\right]}{y_i}} \right),
\end{equation}
which suggests us to define the following pair of affine coordinates,
\begin{equation}
\label{MAP}
Z=\frac{x\left[x^2+\varepsilon y^2\right]}{\varepsilon y}~~\mathrm{and}~~ 
W=\frac{y\left[x^2+\varepsilon y^2\right]}{\varepsilon x} 
\end{equation}

After some cumbersome simplifications we find that the other 
more complicated ratios can
also be expressed in terms of simple 
functions on these new variables. It turns out that 
the transfer matrix eigenvalues can be rewritten as follows,
\begin{eqnarray}
\Lambda(\lambda,\mu)& =& \left[a(\lambda,\mu)\right]^{\mathrm{L}} \prod_{i=1}^{\mathrm{L}-n}
\frac{y}{\varepsilon x} \left( \frac{\varepsilon+\frac{Z_i}{W}}{1-\frac{Z_i}{Z}} \right)
+\left[\overline{b}(\lambda,\mu)\right]^{\mathrm{L}} \prod_{i=1}^{\mathrm{L}-n}
\frac{y}{\varepsilon x} \left( \frac{\varepsilon+\frac{Z_i}{W}}{1-\frac{Z_i}{Z}} \right)
\left(\frac{\frac{1}{\varepsilon}[Z-\frac{\varepsilon}{Z}]-\varepsilon [Z_i -\frac{\varepsilon}{Z_i}]-\mathrm{U}\sqrt{\varepsilon}}
{\varepsilon[Z-\frac{\varepsilon}{Z}]-\frac{1}{\varepsilon} [Z_i -\frac{\varepsilon}{Z_i}]+\mathrm{U}\sqrt{\varepsilon}} \right) \nonumber \\
&+&\left[f(\lambda,\mu)\right]^{\mathrm{L}} \prod_{i=1}^{\mathrm{L}-n}
\frac{y}{x} \left( \frac{\varepsilon+Z Z_i}{W Z_i-1} \right),
\end{eqnarray}
while the Bethe ansatz equations become,
\begin{equation}
\left[\frac{a(\lambda_j,\mu)}{\overline{b}(\lambda_j,\mu)}\right]^{\mathrm{L}}= 
\prod_{\stackrel{i=1}{i \neq j}}^{\mathrm{L}-n}
\frac{\frac{1}{\varepsilon}[Z_j-\frac{\varepsilon}{Z_j}]-\varepsilon [Z_i -\frac{\varepsilon}{Z_i}]-\mathrm{U}\sqrt{\varepsilon}}
{\varepsilon[Z_j-\frac{\varepsilon}{Z_j}]-\frac{1}{\varepsilon} [Z_i -\frac{\varepsilon}{Z_i}]
+\mathrm{U}\sqrt{\varepsilon}},~~ 
j=1,\dots,\mathrm{L}-n.
\end{equation}

Apart from the presence of the second spectral 
parameter $\mu$ associated to the original genus five curve, the main structure of
the eigenvalues and Bethe equations are given in terms of
polynomial ratios depending only on the auxiliary coordinates $Z$ and $W$.
The map (\ref{MAP}) defines a ramified degree four morphism and
from the Hurwitz's formula (see for example \cite{MIRA})
we expect that the image curve $\overline{\mathrm{C}}$ 
depending on the variables $Z$ and $W$ should
have genus less than five. The explicit expression of such curve can be obtained 
eliminating the variables
$x$ and $y$ from Eqs.(\ref{CURVE},\ref{MAP}) and after 
some manipulations we obtain,
\begin{equation}
\overline{\mathrm{C}}= \sqrt{\varepsilon} \left(Z-\frac{\varepsilon}{Z}\right)
+\frac{1}{\sqrt{\varepsilon}} \left(W-\frac{1}{\varepsilon W}\right)+\mathrm{U}=0
\end{equation}
which turns out to be a non-singular cubic curve 
and therefore has genus one.

At this point a natural question to be asked is as follows. Can we choose the 
parameter $\mu$ such that
the eigenvalues structure as well as the Bethe ansatz equations 
become determined basically by
meromorphic functions on these new variables?. By examining the weights (\ref{WEI}) we see that one such 
possibility is to identify the parameter $\mu$ with the point $x=1$ and $y=0$
on the genus five curve (\ref{CURVE}). For this particular choice the weights 
$a(\lambda,\mu)$, $\overline{b}(\lambda,\mu)$ and
$f(\lambda,\mu)$ simplify considerably and the final result for 
the eigenvalues is,
\begin{eqnarray}
\label{EIN1}
\frac{\Lambda(\lambda)}{x^{\mathrm{L}}}& =& \prod_{i=1}^{\mathrm{L}-n}
\frac{y}{\varepsilon x} \left( \frac{\varepsilon+\frac{Z_i}{W}}{1-\frac{Z_i}{Z}} \right)
+\left[\frac{1}{Z}\right]^{\mathrm{L}} \prod_{i=1}^{\mathrm{L}-n}
\frac{y}{\varepsilon x} \left( \frac{\varepsilon+\frac{Z_i}{W}}{1-\frac{Z_i}{Z}} \right)
\left(\frac{\frac{1}{\varepsilon}[Z-\frac{\varepsilon}{Z}]-\varepsilon [Z_i -\frac{\varepsilon}{Z_i}]-\mathrm{U}\sqrt{\varepsilon}}
{\varepsilon[Z-\frac{\varepsilon}{Z}]-\frac{1}{\varepsilon} [Z_i -\frac{\varepsilon}{Z_i}]+\mathrm{U}\sqrt{\varepsilon}} \right) \nonumber \\
&+&\left[\frac{W}{Z}\right]^{\mathrm{L}} \prod_{i=1}^{\mathrm{L}-n}
\frac{y}{x} \left( \frac{\varepsilon+Z Z_i}{W Z_i-1} \right),
\end{eqnarray}
and the respective Bethe ansatz equations are,
\begin{equation}
\label{BET1}
\left[Z_j\right]^{\mathrm{L}}= 
\prod_{\stackrel{i=1}{i \neq j}}^{\mathrm{L}-n}
\frac{\frac{1}{\varepsilon}[Z_j-\frac{\varepsilon}{Z_j}]-\varepsilon [Z_i -\frac{\varepsilon}{Z_i}]-\mathrm{U}\sqrt{\varepsilon}}
{\varepsilon[Z_j-\frac{\varepsilon}{Z_j}]-\frac{1}{\varepsilon} [Z_i -\frac{\varepsilon}{Z_i}]
+\mathrm{U}\sqrt{\varepsilon}},~~ 
j=1,\dots,\mathrm{L}-n.
\end{equation}

From the above results we conclude that apart from a common overall factor $(y/x)^{\mathrm{L}-n}$ 
for the eigenvalues the main important dependence is in fact on the affine variables
of the elliptic curve $\overline{\mathrm{C}}$. It is well known that genus one curves 
degenerate to rational curves when the respective modulus becomes zero. In the case of
the curve $\overline{\mathrm{C}}$ we find that this 
fact occurs, for real $\mathrm{U}$, at the following points,
\begin{equation}
\label{UCRI}
\mathrm{U}_c= \pm 2 \sqrt{3}=\pm 3.464101 \dots.
\end{equation}

We emphasize that the above result does not imply 
that at $\mathrm{U}_c$ the vertex weights 
are necessarily parametrized by rational functions. Although we
expect that at this special coupling the genus five curve is going 
to degenerate into other ones with lower genus they do not need to 
be rational. Indeed, for the coupling values (\ref{UCRI}) the genus
five curve factorizes in terms of the product 
of two cubic curves $\mathrm{C}_{\pm}$. For example in the
case of $\mathrm{U}_c=2 \sqrt{3}$ the
irreducible components are,
\begin{equation}
\mathrm{C}_{\pm}= x^3 \pm \varepsilon x^2y +\varepsilon x y^2 \mp y^3/\varepsilon \mp x +y
\end{equation}
and consequently the weights are sitting in genus one curves.

Next, it is conceivable to think that a variation on the 
geometry of the weights should
be connected with an equivalent change on the physical properties of the vertex
model. This view is somehow based on what happens for instance with the eight-vertex model when it
degenerates into the six-vertex model \cite{BAX}. We stress that our situation here
is different since the number of non-null weights remain the same 
for any coupling $\mathrm{U}$ which brings extra motivation to investigate 
such geometric degeneration. In next section we will study 
this issue in the context of the corresponding spin one chain and we find 
strong evidences that the points $\mathrm{U}_c$ define indeed 
two parameter regions with distinct types of behaviour of the
low-lying excitations.

\section{The spin one chain}
\label{sec3}

The Hamiltonian of the spin chain is derived from Eq.(\ref{HAM}) 
identifying the rapidity $\lambda_0$ 
with the expansion point $x=1$ and $y=0$. We observe that the 
Hamiltonians with positive and negative signs in the exponential
of the root of unity $\varepsilon$  
can be related by means of
a unitary transformation. Here we choose 
$\varepsilon=\exp(\IM \pi/3)$  
and we find that its expression 
is given by,
\begin{eqnarray}
\label{HAM1}
\mathrm{H}(\mathrm{U})&=&\sum_{j=1}^{\mathrm{L}}\Bigg\{-\frac{\exp(-\IM \pi/6)}{2}S_{j}^{+}S_{j+1}^{-}
-\frac{\exp(\IM \pi/6)}{2}S_{j}^{-}S_{j+1}^{+}
-\frac{\IM}{2} \left(S_{j}^{+}S_j^{z}\right) S_{j+1}^{-}
+\frac{\IM}{2} S_{j}^{-}\left(S_{j+1}^{+} S_{j+1}^{z}\right)  \nonumber \\
&+&\frac{\mathrm{U}}{2}\left(S_j^{z}\right)^2 \Bigg\}
\end{eqnarray}
where $S_{j}^{+},S_{j}^{-}$ and $S_j^{z}$ are the
spin-$1$ matrices generators of the
$\mathrm{SU}(2)$ algebra,
\EQ
\mathrm{S}_{j}^{+}=\sqrt{2}\left[
\begin{array}{ccc}
0 & 1 &0 \\
0 & 0 &1 \\
0 & 0 &0 \\
\end{array}
\right]_{j},~~
\mathrm{S}_{j}^{-}=\sqrt{2}\left[
\begin{array}{ccc}
0 & 0 &0 \\
1 & 0 &0 \\
0 & 1 &0 \\
\end{array}
\right]_{j},~~
\mathrm{S}_{j}^{z}=\left[
\begin{array}{ccc}
1 & 0 &0 \\
0 & 0 &0 \\
0 & 0 &-1 \\
\end{array}
\right]_{j},
\EN

The spectrum of the Hamiltonian (\ref{HAM1}) can be determined for each sector
of the magnetization directly from 
the transfer matrix Bethe 
ansatz solution described in previous section.
By setting
the variable $Z_j=\exp(\IM k_j)$ in Eq.(\ref{BET1}), where $k_j$ 
plays the role of quasiparticle momenta, we find that they are
constrained by the following
equations\footnote{The diagonalization of this Hamiltonian
can be also done along the lines of the coordinate Bethe 
ansatz method see for instance \cite{CAMP}.},
\begin{equation}
\label{BET2}
\exp(\IM k_j \mathrm{L})= \prod_{i=1}^{\mathrm{L}-n} \frac{\sin(k_j-\pi/6)/\exp(\IM \pi/3) -\sin(k_i-\pi/6)\exp(\IM \pi/3)+\IM\mathrm{U}/2}
{\sin(k_j-\pi/6)\exp(\IM \pi/3) -\sin(k_i-\pi/6)/\exp(\IM \pi/3)-\IM \mathrm{U}/{2}},~~j=1,\dots,\mathrm{L}-n,
\end{equation}

The respective eigenenergies are obtained by taking the 
logarithmic derivative of Eq.(\ref{EIN1}) at
the expansion point. As a result we obtain,
\begin{equation}
\label{energia}
\mathrm{E}_n(\mathrm{U})= -\sum_{j=1}^{\mathrm{L}-n} 2 \cos(k_j+\pi/6) +n \frac{\mathrm{U}}{2}.
\end{equation}

In the case of even lattice sizes the eigenvalues of the Hamiltonian
for positive and negative values of the parameter $\mathrm{U}$ can be related.
We can for example rotate every spin at the odd sites by an angle $\pi$ about the
azimuthal direction and this unitary transformation implies,
\begin{equation}
\label{SIM}
\mathrm{H}(\mathrm{U})=-\mathrm{H}(-\mathrm{U})
\end{equation}
%stablishing a correspondence between the {\bf high-energy} eigenstates for $\mathrm{U} >0$ with the 
%{\bf low-energy} eigenstates for $\mathrm{U}<0$. 

In order to gain some insight about the spectrum of 
Hamiltonian (\ref{HAM1}) we have performed
exact diagonalization for lattice sizes $ \mathrm{L} \leq 12$. 
Although the Hamiltonian 
is non-Hermitian we find that the lowest eigenvalues in any sector $n$ 
are always real numbers for generic values of $\mathrm{U}$.
The imaginary eigenvalues occur in
complex conjugated pairs and as $\mathrm{L}$ grows we have observed  
that their real parts do not appear to contribute to the low-lying 
states in the eigenspectrum.
In addition to that, we note that for
a given size $\mathrm{L}$ there exits a critical coupling $\mathrm{U}(\mathrm{L})$
such that for $\mathrm{U} \geq \mathrm{U}(\mathrm{L})$ all the Hamiltonian eigenvalues are
real numbers. This analysis has been performed up to $\mathrm{L}=9$ because 
it requires the knowledge of the full spectrum and our findings are exhibited in 
Table (\ref{tab1}).
\begin{table}[h]
\begin{center}
\begin{tabular}{|r|l|l|l|l|l|l|} \hline 
$\mathrm{L}$ & 4 & 5 & 6 & 7 & 8 & 9  \\ \hline \hline
$\mathrm{U}(\mathrm{L})$ & 2.99684 & 3.1637 & 3.25492 & 3.3101 & 3.34601 & 3.3707 \\ \hline 
\end{tabular}
\end{center}
\caption{The Hamiltonian has entirely real energy spectrum for $\mathrm{U} \geq \mathrm{U}(\mathrm{L})$.} 
\label{tab1}
\end{table}

\begin{figure}[ht]
\begin{center}    
\includegraphics[width=9cm]{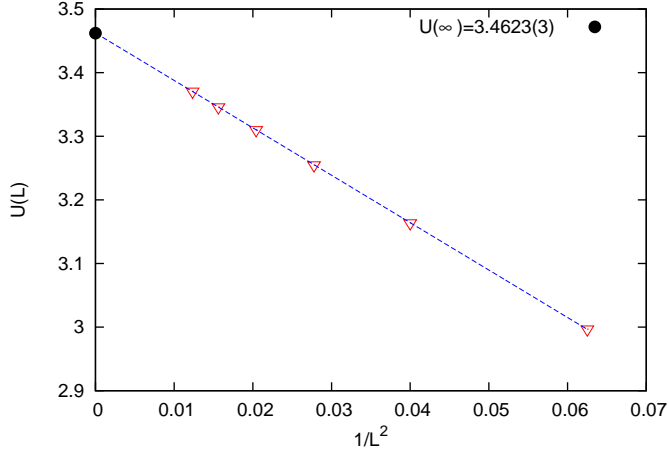}
\end{center}
\caption{The fitting of Table (\ref{tab1}) with $\mathrm{L}^{-2}$ and the extrapolated $\mathrm{U}(\infty)$ value.}
\label{fig1}
\end{figure}

The slow grow of $\mathrm{U}(\mathrm{L})$ with the length
$\mathrm{L}$ indicates that it will extrapolate
to a finite number for large sizes. We search for a linear fitting of the finite-size data 
and find that its dependence on the length
appears to be dominated by the power $1/\mathrm{L}^2$. This is shown in Figure (\ref{fig1})
and remarkably the 
extrapolated value $\mathrm{U}(\infty)$ is very close to 
the critical point (\ref{UCRI})
uncovered via geometrical considerations. While is tempting to conjecture that
$\mathrm{U}(\infty)=2\sqrt{3}$ 
an analytical explanation for this coincidence has eluded us so far.

\subsection{The regime $\mathrm{U} \geq 0$}

\begin{figure}[ht]
\begin{minipage}{0.5\linewidth}
\begin{center}    
\includegraphics[width=9cm]{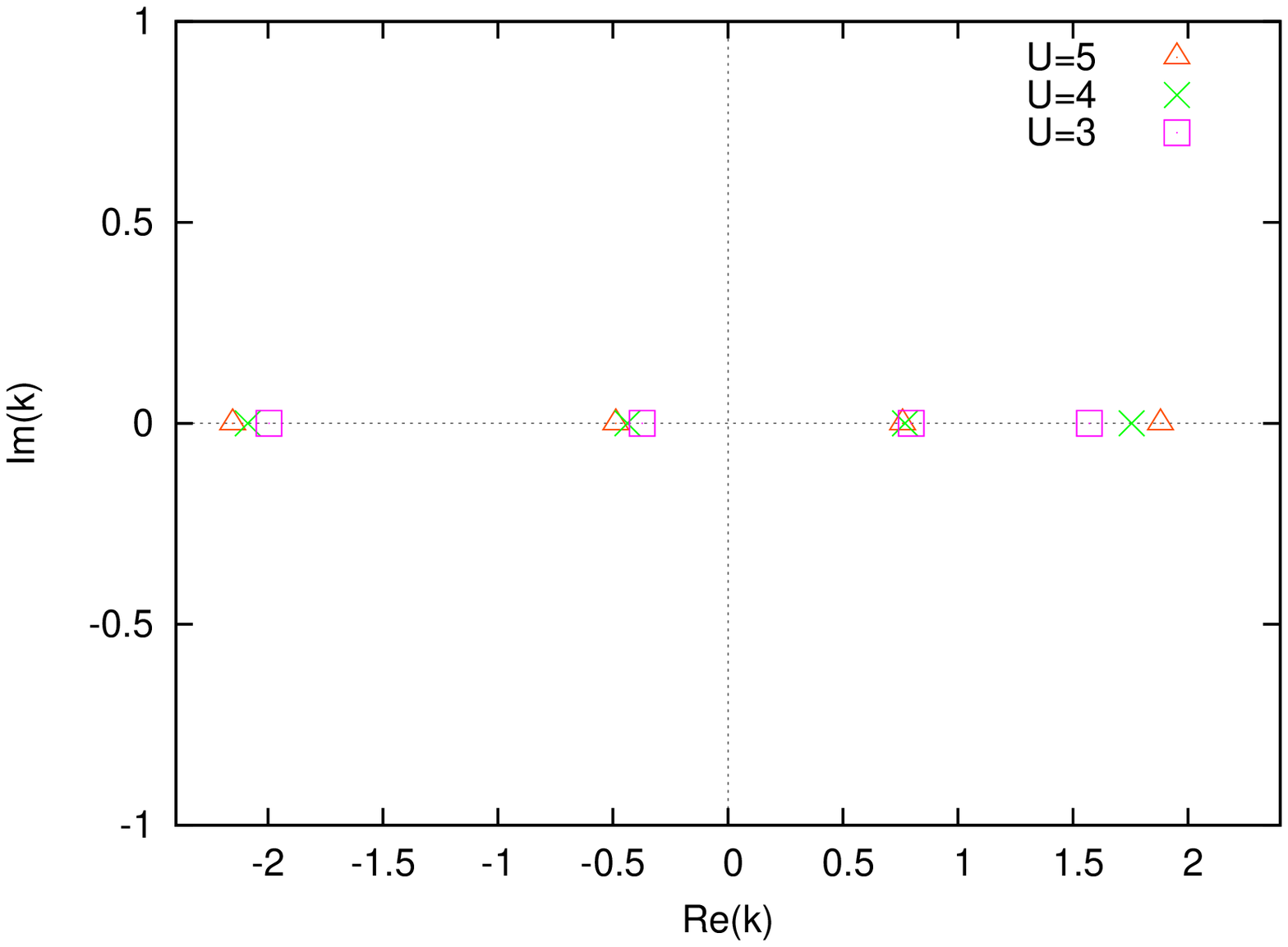}
\end{center}
\end{minipage}
\begin{minipage}{0.5\linewidth}
\begin{center}    
\includegraphics[width=9cm]{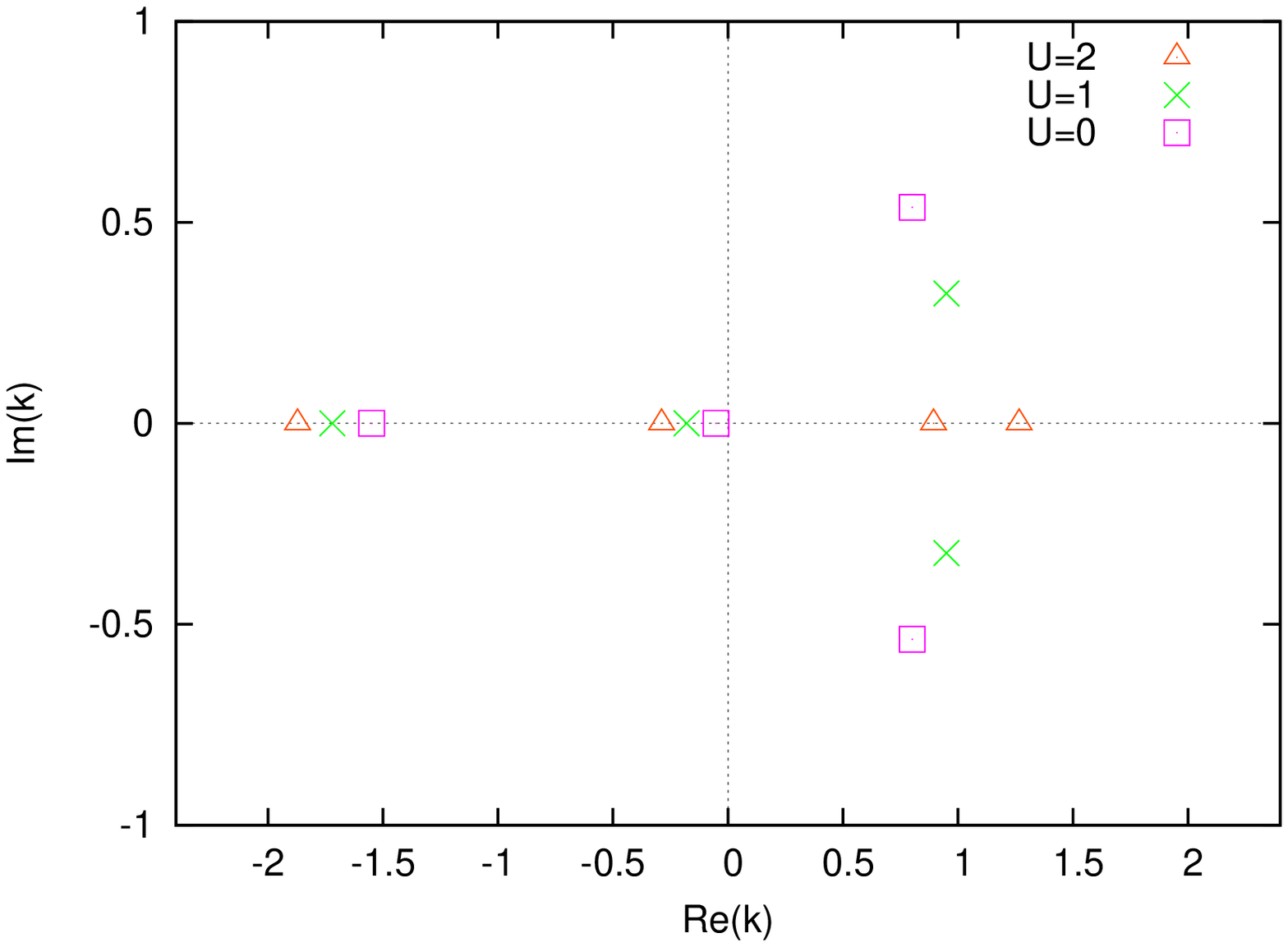}
\end{center}
\end{minipage}
\caption{Ground state Bethe roots for $\mathrm{L}=4$ and for some values of $\mathrm{U}\geq 0$. }
\label{fig2}
\end{figure}

\begin{figure}[ht]
\begin{minipage}{0.5\linewidth}
\begin{center}    
\includegraphics[width=9cm]{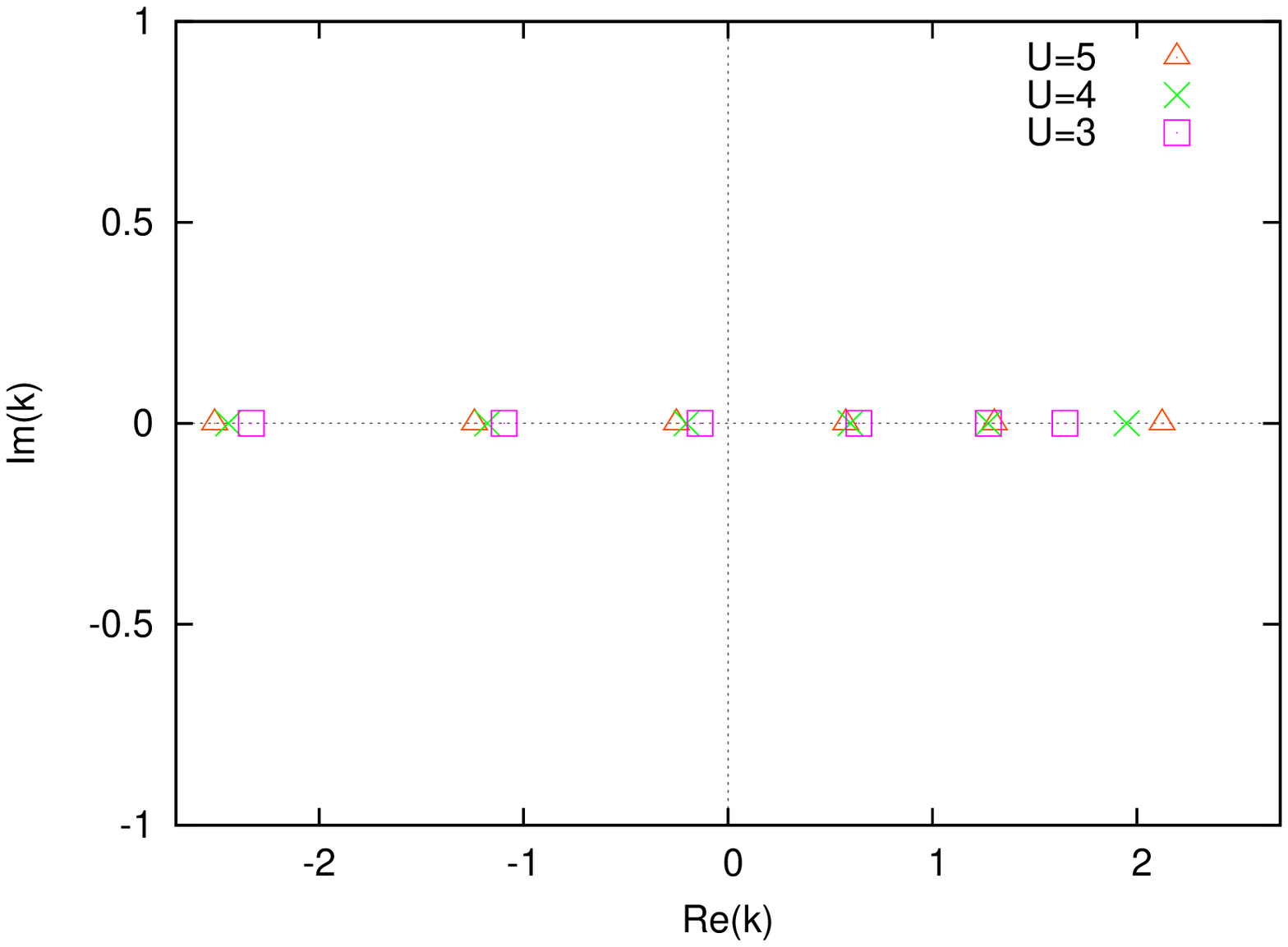}
\end{center}
\end{minipage}
\begin{minipage}{0.5\linewidth}
\begin{center}    
\includegraphics[width=9cm]{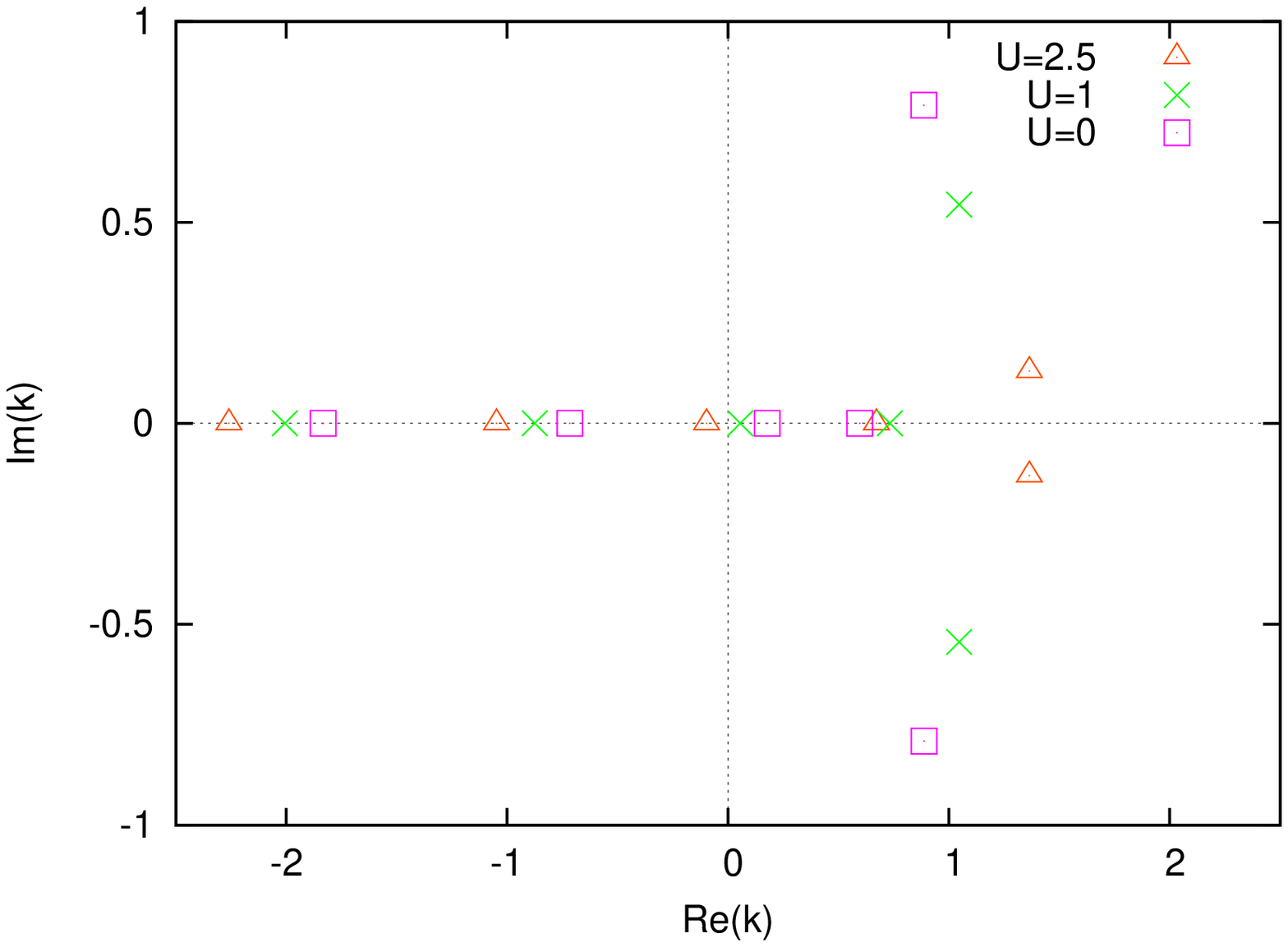}
\end{center}
\end{minipage}
\caption{Ground state
Bethe roots for $\mathrm{L}=6$ and for some values of $\mathrm{U} \geq 0$. }
\label{fig3}
\end{figure}

We start our analysis by solving numerically the Bethe 
ansatz equation (\ref{BET2}) for small sizes in order to 
find the structure of the roots
distribution for low-lying states.  As expected we find that the ground state 
sits in the sector with zero magnetization $n=0$.
In Figures (\ref{fig2},\ref{fig3}) we exhibit the ground 
state roots in the complex plane for $\mathrm{L}=4,6$.

We note that the Bethe roots are real until we 
reach some 
value of $\mathrm{U}$ and after this critical point 
a pair of real zeros combine together to form 
a string of length two. As we increase the lattice size
such critical point appears to grow slowly 
towards an irrational 
number which we conjecture to be exactly
$\mathrm{U}_c=2 \sqrt{3}$. In order to provide some support
to this expectation consider 
the specific example of a state at the sector $n=\mathrm{L}-2$ 
built out of the complex pair $k_{1}=\alpha -\IM \beta$ 
and $k_{2}=\alpha +\IM \beta$. We can assume 
$\beta >0$ without any loss of generality. For large sizes
the left hand side of 
the Bethe equation (\ref{BET2}) for $k_1$ grows 
exponentially with 
$\mathrm{L}$. The only way to fulfill this behaviour
is that the respective right hand side 
becomes exponentially 
close to a pole, namely
\begin{equation}
\label{bound}
\sin(k_1-\pi/6)\exp(\IM \pi/3) =\sin(k_2-\pi/6)/\exp(\IM \pi/3)+\IM \mathrm{U}/{2}+\mathcal{O}\left[\exp(-\mathrm{L})\right]
\end{equation}

By the same token, for the root $k_2$ we expect that the 
right hand side goes to zero which leads us to 
the same condition (\ref{bound}). From our numerical
solution we see that
near to the critical point the 
imaginary part of the roots is  very 
small and therefore the trigonometric functions
should be limited by a unity. Taking this fact into 
account in Eq.(\ref{bound}) we
obtain the following bound for the coupling, 
\begin{equation}
\mathrm{U} \leq 2 \left[\frac{\exp(\IM \pi/3) -\exp(-\IM \pi/3)}{\IM} \right]= 2\sqrt{3}
\end{equation}

\begin{figure}[ht]
\begin{minipage}{0.5\linewidth}
\begin{center}    
\includegraphics[width=9cm]{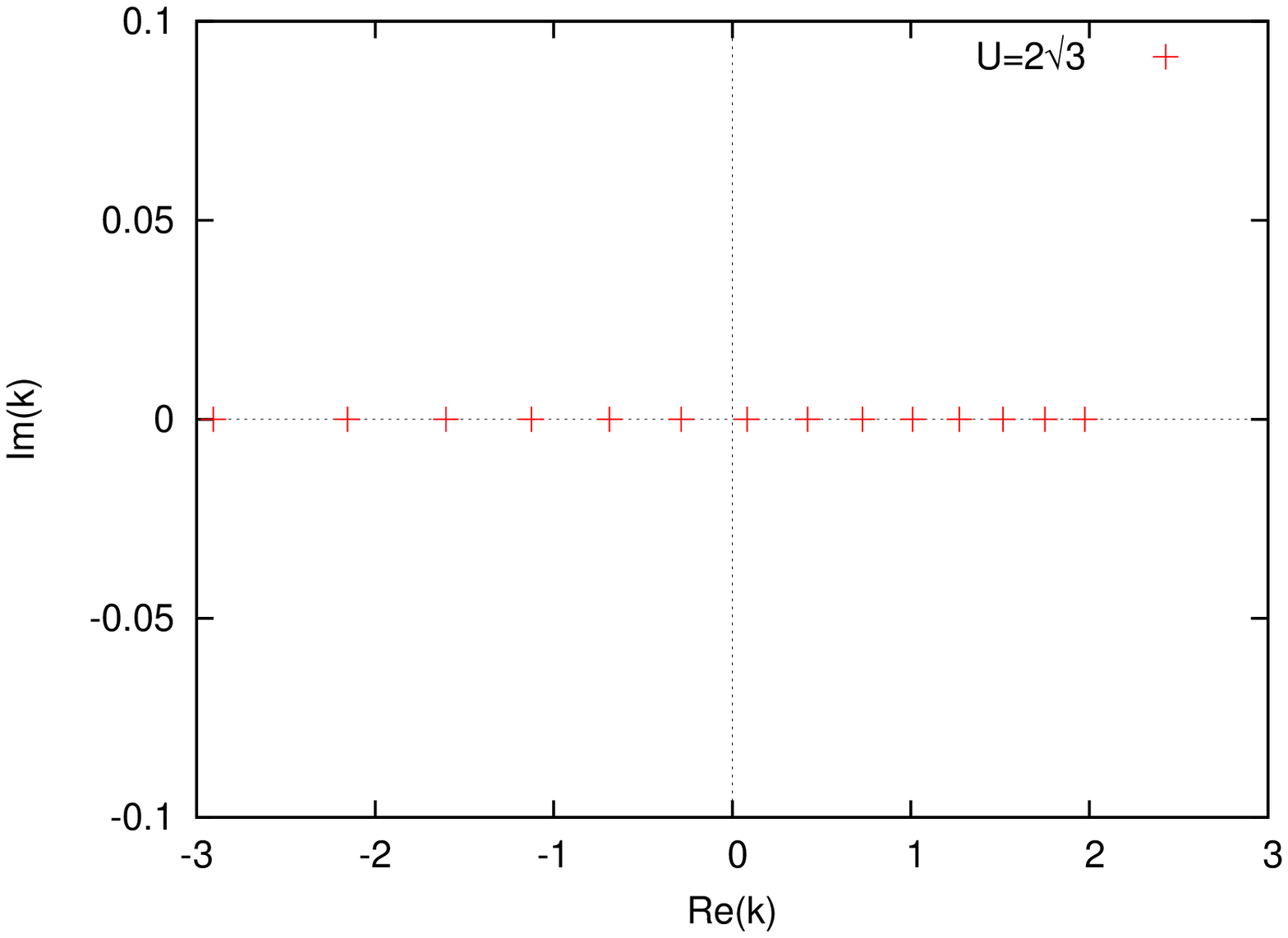}
\includegraphics[width=9cm]{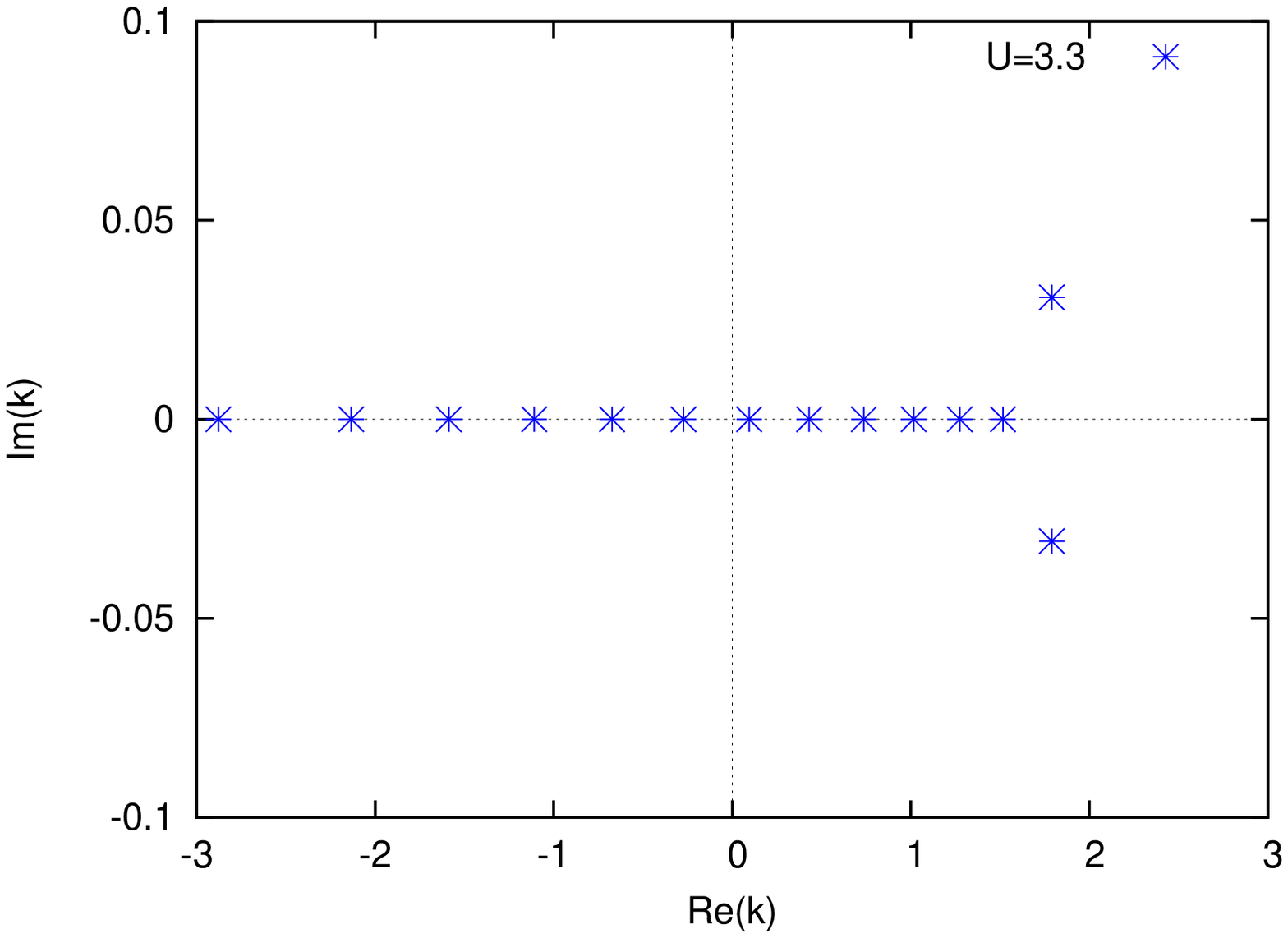}
\end{center}
\end{minipage}
\begin{minipage}{0.5\linewidth}
\begin{center}    
\includegraphics[width=9cm]{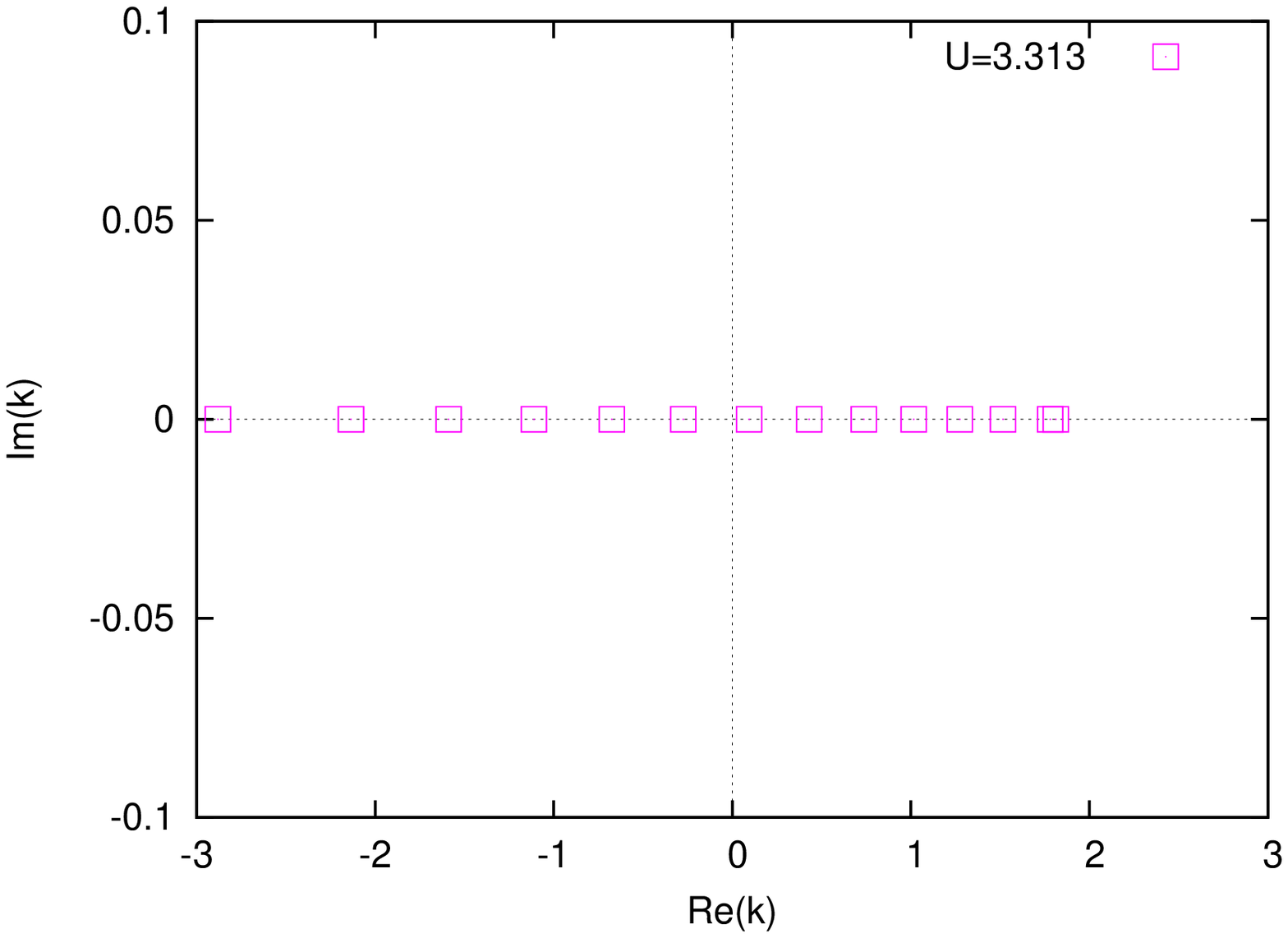}
\includegraphics[width=9cm]{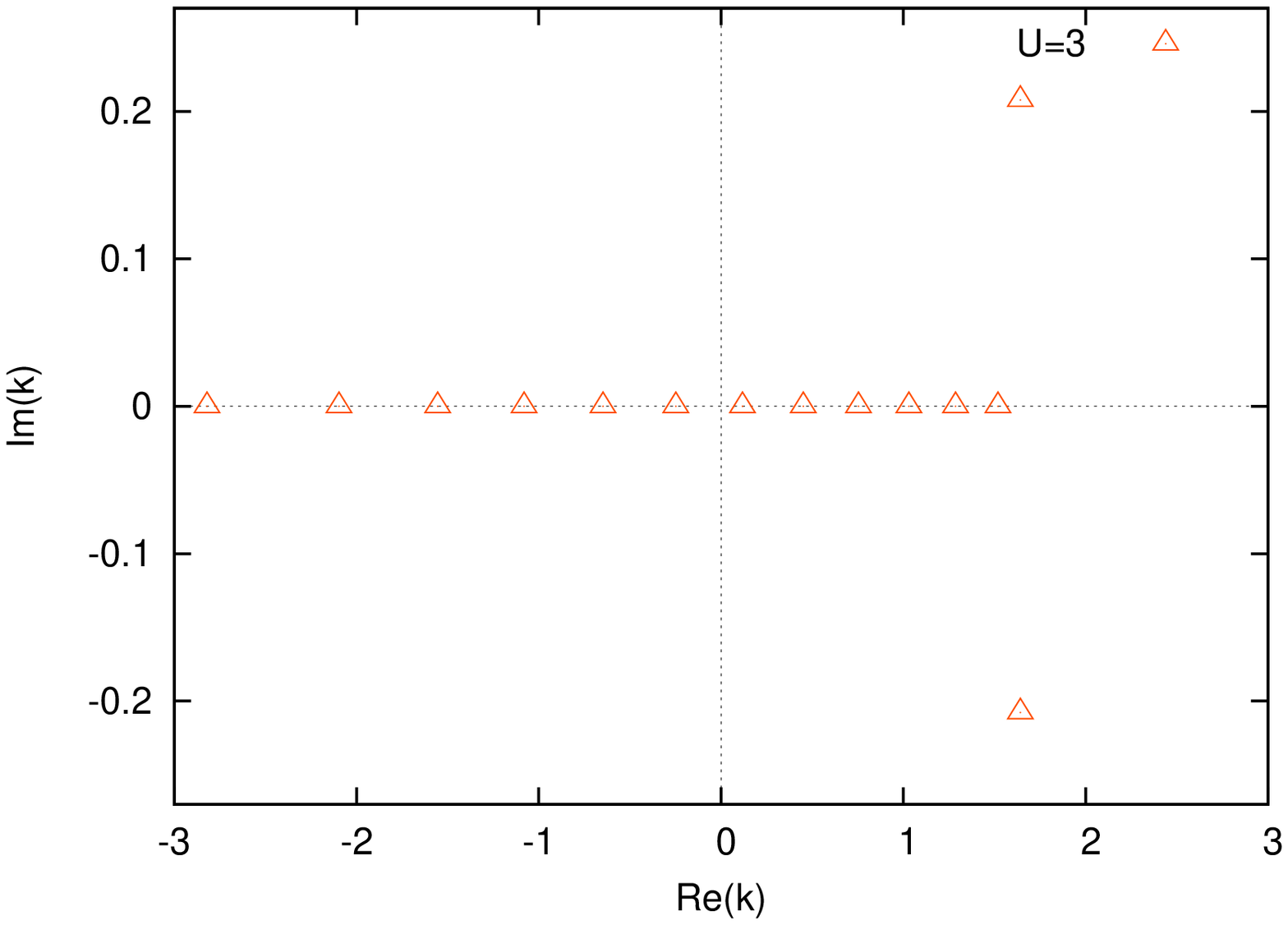}
\end{center}
\end{minipage}
\caption{Ground state Bethe roots for $\mathrm{L}=14$ and for some values of $\mathrm{U}$ close to $\mathrm{U}_c$.}
\label{fig4}
\end{figure}

Another piece of evidence comes from our 
numerical solution 
of the Bethe equations for large lattice sizes. We have 
verified that 
real roots are indeed stable solutions in the range  
$\mathrm{U} \geq 2 \sqrt{3}$
up to $\mathrm{L}=1024$ sites. Now suppose  
we have at hand the real roots 
solution at $\mathrm{U}_c$ for some large enough lattice size. 
When we try to iterate it in order 
to obtain the solution 
on the neighborhood $\mathrm{U}=2\sqrt{3}-\epsilon$ we find out that 
the first two largest positive roots get very close to each other
even for small $\epsilon$. By increasing $\epsilon$,
instead of having coincident real roots,
we noticed that such pair of roots will give rise to a string with
a small non-zero imaginary part. This scenario can be seen already
for $\mathrm{L}=14$ as illustrated in Figures (\ref{fig4}).

The dependence of the Bethe roots configurations on 
the coupling $\mathrm{U}$ tells us
that the low-energy properties of the spin chain have to be studied separately
in the two distinct intervals: $\mathrm{U} \geq 2\sqrt{3}$ and
$0\leq \mathrm{U} <2\sqrt{3}$. We shall see below that the spin chain has a gap
for first range of the coupling while appears to be massless on the second one.

\subsubsection{The range $\mathrm{U} \geq 2\sqrt{3}$}

In this case we have found that the scenario described above 
for the ground state remains
valid for the lowest state in each sector $n$. These states 
are all described by real solutions to the Bethe equations. Therefore we can
take the logarithm of Eq.(\ref{BET2}) and as result we obtain,
\begin{equation}
\label{BET3}
\mathrm{L} k_j=2 \pi Q_j+2\sum_{i=1}^{\mathrm{L}-n} 
\arctan\left[ \frac{\sin(k_j-\pi/6)-\sin(k_i-\pi/6)}{
\sqrt{3}\left[\sin(k_j-\pi/6)+\sin(k_i-\pi/6)\right]-\mathrm{U}}\right]
,~~j=1,\dots,\mathrm{L}-n,
\end{equation}
where the numbers $Q_j$ define the different branches of the logarithm
and they have to be chosen integer
or half-integer according to the rule,
\begin{equation}
\label{QN}
Q_j= \frac{\mathrm{L}-n-1}{2}-(j-1),~~j=1,\dots,\mathrm{L}-n.
\end{equation}

In the thermodynamic limit the solutions $k_j$ 
become continuously distributed on some interval
$k_{min} \leq k \leq k_{max}$ with a density of roots $\sigma(k)$.
By taking the difference between next
neighbors roots satisfying Eq.(\ref{BET3}) one gets 
an integral equation for such density function which is,
\begin{equation}
\label{INT}
2\pi \sigma(k)=1 +2\cos(k-\pi/6)\int_{k_{min}}^{k_{max}} \frac{
\left[\mathrm{U}+ \sqrt{3} \mathrm{F}_{-}(k,k^{'})-\sqrt{3}\mathrm{F}_{+}(k,k^{'}) \right]}{
[\mathrm{F}_{-}(k,k^{'})]^2+[\mathrm{U}-\sqrt{3}\mathrm{F}_{+}(k,k^{'})]^2} \sigma(k^{'}) d k^{'},
\end{equation}
where the kernel functions are defined by,
\begin{equation}
\mathrm{F}_{\pm}(x,y)=\sin(x-\pi/6) \pm \sin(y-\pi/6).
\end{equation}

The integrate density gives the total number of roots per site which
is fixed by the range of integration. For the ground state we have
the constraint, 
\begin{equation}
\label{constra}
\int_{k_{min}}^{k_{max}} \sigma(k) dk=1.
\end{equation}

\begin{figure}[ht]
\begin{center}    
\includegraphics[width=9cm]{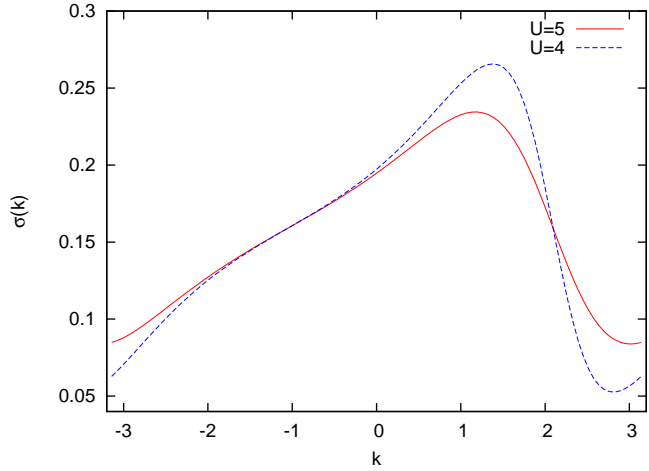}
\end{center}
\caption{The root density $\sigma(k)$ for $\mathrm{U}=5$ and $\mathrm{U}=4$.}
\label{fig5}
\end{figure}

We have solved Eqs.(\ref{INT}) by iterative numerical
integration and found out that the condition (\ref{constra}) 
is satisfied provide we set 
$[k_{min},k_{max}]=[k_0,k_0+2\pi]$ where $k_0 \in \mathbb{R}$ is
an arbitrary integration point. 
%This can be analytically checked 
%by first integrating Eq.(\ref{INT}) on this interval and next
%showing that the right hand side double integral is zero with
%the help of methods of contour integration. 
In Figure (\ref{fig5})
we have plotted the pattern of $\sigma(k)$ on a symmetrical
interval with the choice $k_0=-\pi$.

The expression for the ground state energy per site 
$e_0(\mathrm{U})$ is derived from
Eq.(\ref{energia}). In the thermodynamic limit 
$\mathrm{L} \rightarrow \infty$ we obtain,
\begin{equation}
\label{BULK}
e_0(\mathrm{U})=
-2\int_{k_0}^{k_0+2\pi} \cos(k+\pi/6) \sigma(k) dk
\end{equation}

In order to illustrate the ground state 
behaviour we present 
in Table (\ref{tab2}) the respective 
energy for finite 
sizes together with bulk value obtained from the numerical
solution of the integral equation (\ref{INT}).
\begin{table}[ht]
\begin{center}
\begin{tabular}{|c||c|c|c|c|}
  \hline
$\mathrm{E}_0(\mathrm{U})/\mathrm{L}$ & $\mathrm{U}=5$ & $\mathrm{U}=4.5$ & $\mathrm{U}=4$ & $ \mathrm{U}=2\sqrt{3}$ \\ \hline \hline
 8  & -0.2044~6422~8606   & -0.2300~4351~1273 & -0.2636~1069~8228 &   -0.3118~1225~6414   \\
12 & -0.2012~6573~7835   & -0.2248~9993~6711 &  -0.2560~4511~7699 &   -0.3024~9083~5111    \\
16 & -0.2008~2406~1600   & -0.2238~7105~9808 &  -0.2538~9742~0183 &   -0.2992~6241~0687    \\
24 & -0.2007~3651~4467   & -0.2235~5132~5431 &  -0.2528~0569~9816 &   -0.2969~6872~3330    \\
64 & -0.2007~3305~6600   & -0.2235~2248~8844 &  -0.2525~4016~7269 &   -0.2953~9909~9678  \\
128 & -0.2007~3305~6598  & -0.2235~2248~8276 &  -0.2525~3979~7503 &   -0.2952~0705~3381    \\
256 & -0.2007~3305~6598  & -0.2235~2248~8276 &  -0.2525~3979~7464 &   -0.2951~5908~1267   \\
516 & -0.2007~3305~6598  & -0.2235~2248~8276 &  -0.2525~3979~7464 &   -0.2951~4703~1065  \\
1024 & -0.2007~3305~6598 & -0.2235~2248~8276 &  -0.2525~3979~7464 &   -0.2951~4409~6192  \\
$e_{0}(\mathrm{U})$ & -0.2007~3305~6598  & -0.2235~2248~8276 & -0.2525~3979~7464  & -0.2951~4306~683   \\ \hline
\end{tabular}
\caption{Finite-size sequences for the ground state energy per site and the respective bulk values.}
\label{tab2}
\end{center}
\end{table}

We observe that for first three parameter
values the finite size corrections to the energy become 
irrelevant already after few hundred sites 
typical of an exponential convergence towards 
the thermodynamic
limit. This is an indication that the
low energy physics of the spin chain should be governed by
massive excitations whose gap
appears to decrease as we approach the 
critical point $\mathrm{U}_c=2\sqrt{3}$.
In order to substantiate this expectation we have to 
compute the energy needed to
remove one particle from the ground state. The removal of one root
implies that the momenta $k_j^{'}$ in the sector $n=1$ are 
shifted with respect to
the ground state by small amounts,
\begin{equation}
k_j^{'} = k_j +\frac{\delta(k_j)}{\mathrm{L}}
\end{equation}
where $k_j$ denotes the ground state momenta. 

We now take the difference of Eq.(\ref{BET3}) for the momenta $k_j^{'}$ and $k_j$ and
keep the leading term on $\delta(k_j)$. Replacing the sums by integrals and making use of
Eq.(\ref{INT}) we are led to the following integral equation,
\begin{equation}
\label{INT1}
2\pi \rho(k)=\int_{k_{min}}^{k_{max}} \cos(k^{'}-\pi/6) \frac{
\left[\mathrm{U}+ \sqrt{3} \mathrm{F}_{-}(k,k^{'})+\sqrt{3}\mathrm{F}_{+}(k,k^{'}) \right]}{
[\mathrm{F}_{-}(k,k^{'})]^2+[\mathrm{U}-\sqrt{3}\mathrm{F}_{+}(k,k^{'})]^2} \rho(k^{'}) d k^{'},
\end{equation}
where $\rho(k)=\delta(k) \sigma(k)$.

The lowest energy gap $\Delta(\mathrm{U})$  is computed with the help of Eq.(\ref{energia}), namely
\begin{eqnarray}
\label{GAP}
\Delta(\mathrm{U})&=& \mathrm{E}_1(\mathrm{U})-\mathrm{E}_0(\mathrm{U})= \frac{\mathrm{U}}{2}
-2\sum_{j=1}^{\mathrm{L}-1} \cos(k_j^{'}+\pi/6) 
+2\sum_{j=1}^{\mathrm{L}} \cos(k_j+\pi/6) \nonumber \\
&=& \frac{\mathrm{U}}{2} -2 \cos(\pi/6)
-2\sum_{j=1}^{\mathrm{L}-1} \left[\cos(k_j^{'}+\pi/6) 
-\cos(k_j+\pi/6) \right] \nonumber \\
&=& \frac{\mathrm{U}}{2} -\sqrt{3} +2 \int_{k_{min}}^{k_{max}} \sin(k+\pi/6) \rho(k)
\end{eqnarray}

We have solved the integral equation (\ref{INT1}) numerically for several values 
$\mathrm{U} > 2 \sqrt{3}$. 
We find that the density $\rho(k)$ is very small in the entire momenta interval 
and it seems reasonable to assume that this 
function should vanish.
Taking into account this hypothesis 
the third term in Eq.(\ref{GAP})
is zero and we conjecture that the gap is given by,
\begin{equation}
\label{GAP1}
\Delta(\mathrm{U})= \frac{\mathrm{U}}{2}-\sqrt{3}.
\end{equation}

In order to substantiate the above conclusion we have presented in Table (\ref{tab3}) 
the finite-size sequences for the energy gap. We note that the numerical value around
hundred sites is already very close to the exact prediction (\ref{GAP1}).
\begin{table}[ht]
\begin{center}
\begin{tabular}{|c||c|c|c|c|}
  \hline
$\Delta(\mathrm{U})$ & $\mathrm{U}=5$ & $\mathrm{U}=4.5$ & $\mathrm{U}=4$ & $ \mathrm{U}=2\sqrt{3}$ \\ \hline \hline
4 & 1.0363~1964~4464   &  0.8545~8007~1405 &  0.6921~7452~4898  &  0.5439~9429~5694    \\
6 & 0.8733~4017~7602   &  0.6748~6896~6228 &  0.5023~06541~1977  &  0.3556~2898~7118    \\
8 & 0.8136~3837~7809  &  0.5996~3177~1038 &  0.4144~2122~0768  &  0.2648~5127~5979    \\
10 & 0.7886~6944~1708 &   0.5627~0418~0977 & 0.3652~8046~2750  &  0.2111~5429~3739    \\
12 & 0.7775~9180~8258 &   0.5431~8460~2057 &  0.3349~8712~9854  &  0.1756~15006~7247   \\
24 & 0.7680~7368~9898 &   0.5189~8852~1734  &  0.2776~5781~2737 & 0.0874~7009~7673   \\
64 &  0.7679~4919~2567 &   0.5179~4924~6894  & 0.2679~8469~7157 & 0.0327~4775~5751    \\
128 & 0.7679~4919~2431  &   0.5179~4919~2431  & 0.2679~4919~9834 & 0.01636~7736~1663   \\
Conj. & 0.7679~4919~2431 &  0.5179~4919~2431  & 0.2679~4919~2431  &  0. \\ \hline
\end{tabular}
\caption{ Finite-size sequences for the lowest energy gap and the conjectured value (\ref{GAP1}).}
\label{tab3}
\end{center}
\end{table}

\subsubsection{The range $0 \leq \mathrm{U} < 2\sqrt{3}$}

In this situation we find that the numerical solution of the 
Bethe equations becomes very sensible 
with the initial guess for the roots. This makes it difficult to have 
an idea how the real roots
bound together to form complex strings as we increase the lattice size.   
Besides the presence of real  
and two-string type of roots we can not rule out
the formation of more involved bounds such as strings with multiple imaginary roots.
As consequence of that we have not been able to figure out the
formulation of the string hypothesis in the thermodynamic limit
for this regime of the coupling. Here we have resorted to the results
of the exact diagonalization of the Hamiltonian. In Table (\ref{tab4}) we
exhibit the finite size estimates for the mass gap associated to 
the first excited state. The extrapolated data shows that we have
a very small gap for several values of the coupling suggesting that 
this phase should be governed by massless excitations.
\begin{table}[ht]
\begin{center}
\begin{tabular}{|c||c|c|c|c|}
  \hline
$\Delta(\mathrm{U})$ & $\mathrm{U}=3$ & $\mathrm{U}=2$ & $\mathrm{U}=1$ & $ \mathrm{U}=0$ \\ \hline \hline
4 &  0.4400~1651~0622  & 0.2935~3529~0332  &   0.2295~6105~6843   & 0.2067~1526~6807  \\
5 &  0.3331~8629~3755  & 0.2147~7955~3874  &   0.1739~5124~9866  &  0.1566~0837~5958  \\
6 &  0.2650~0019~6383  & 0.1685~2500~3057  &   0.1423~1283~0329  &  0.1309~7377~8878  \\
7 &  0.2177~9262~8767  & 0.1386~4806~2734  &   0.1211~3457~3771  &  0.1117~7810~9994  \\
8 &  0.1833~3776~7402  & 0.1180~8109~0063  &   0.1058~6440~2313 &   0.0978~6465~8974  \\
9 &  0.1572~1529~3302  & 0.1031~7113~4329  &   0.0941~9346~0366  &  0.0869~9839~9570  \\
10 & 0.1368~3356~0153  & 0.0918~9277~3383  &   0.0849~2246~3102
  &  0.0783~3772~5112  \\
11 & 0.1205~6941~4216  & 0.0830~5359~1425  &   0.0773~4872~5737  &  0.0712~4801~5197  \\
12 & 0.1073~5348~9189  & 0.0759~1868~2286  &   0.0710~3084~7203  &  0.0653~4014~1516  \\
Extrap. & 0.0125($\pm 2$)  & 0.00095 ($\pm 1$)   & 0.00072 ($\pm 2$)  & 0.00057 ($\pm 1$)  \\ \hline
\end{tabular}
\caption{ Finite-size sequences for the energy gap $\mathrm{E}_1(\mathrm{U})-\mathrm{E}_0(\mathrm{U})$ and the extrapolated value for $0 \leq \mathrm{U} < 2\sqrt{3}$.}
\label{tab4}
\end{center}
\end{table}

\subsection{The regime $\mathrm{U} \leq 0$}

\begin{figure}[ht]
\begin{minipage}{0.5\linewidth}
\begin{center}    
\includegraphics[width=9cm]{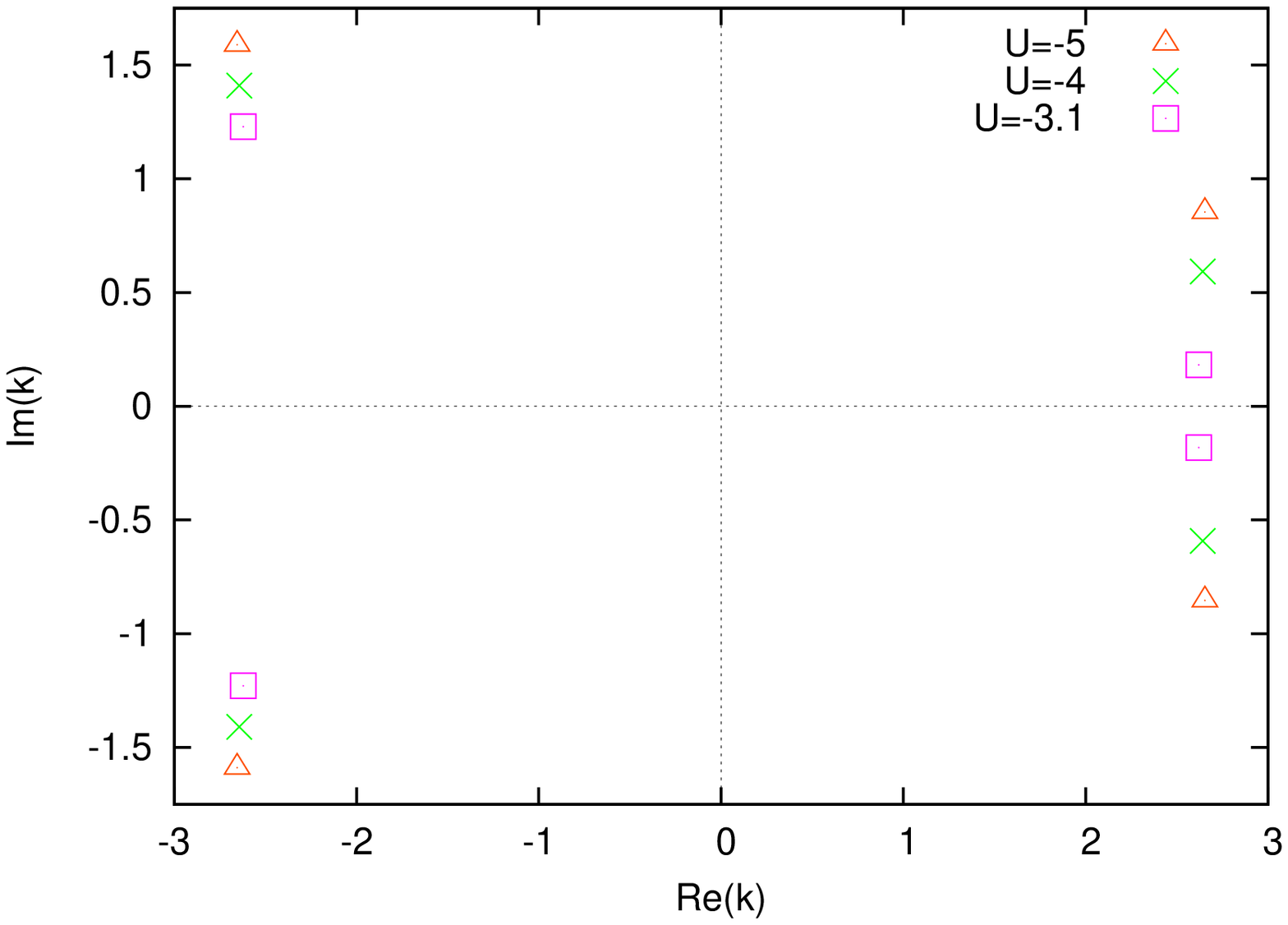}
\end{center}
\end{minipage}
\begin{minipage}{0.5\linewidth}
\begin{center}    
\includegraphics[width=9cm]{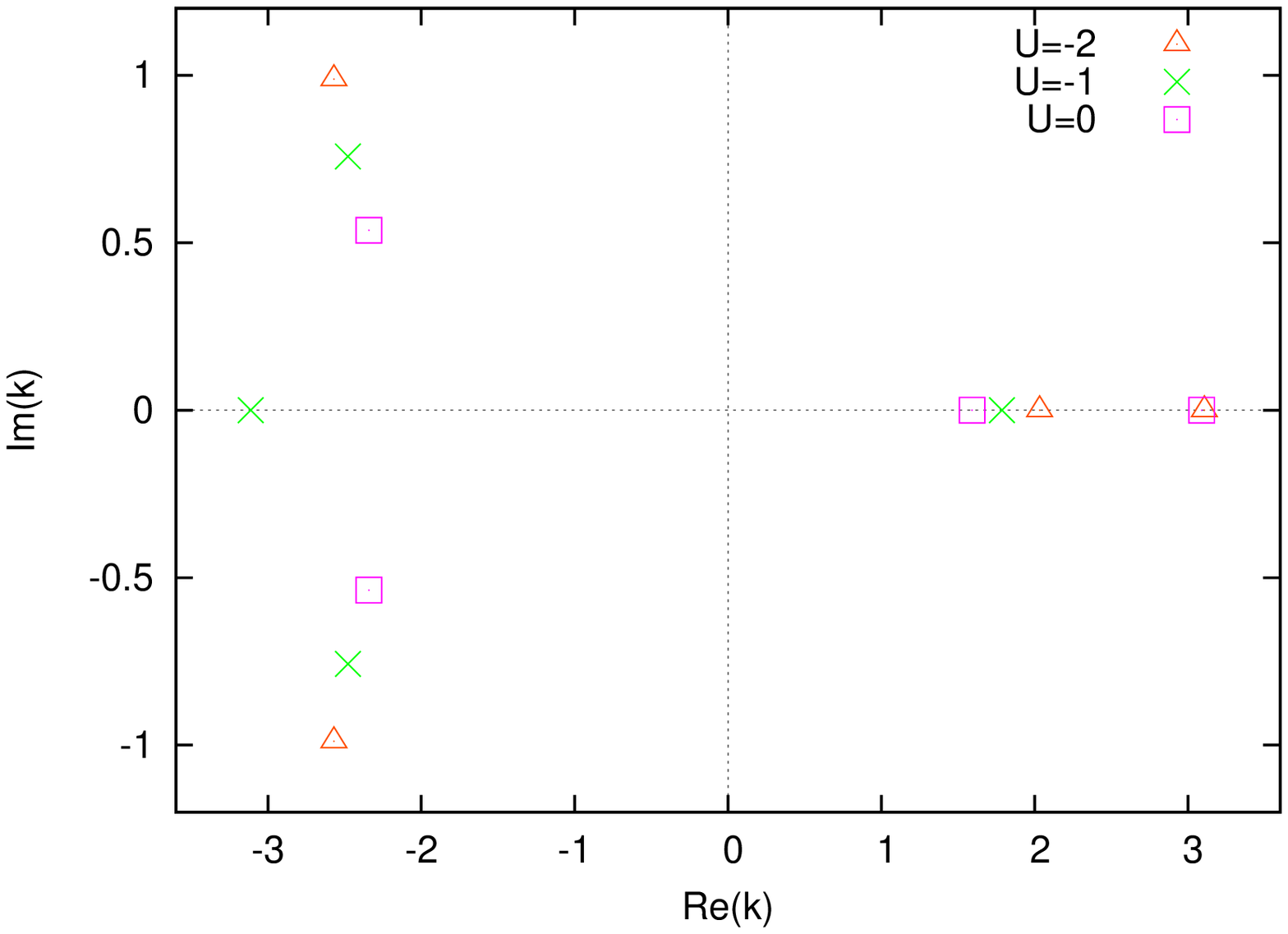}
\end{center}
\end{minipage}
\caption{Ground state Bethe roots for $\mathrm{L}=4$ and 
for some values of  $\mathrm{U}\leq 0$.}
\label{fig6}
\end{figure}

We start the analysis by solving the Bethe equations for small sizes. In Figure (\ref{fig6})  we show
the pattern of the roots for the ground state with $\mathrm{L=4}$. We note that there is a region of the coupling
dominated two types of two-strings
roots having different imaginary parts of the form $-a \pm \IM b$ and $a \pm \IM c$ such that $b-c \sim \mathrm{U}/2$.
Once we decrease the value of $\mathrm{U}$ the complex pair with the lowest imaginary part breaks down
giving rise to two real roots. It turns out that for negative $\mathrm{U}$ we find it difficult 
to obtain the solution 
of the Bethe equations
even for small sizes. Again
this fact has prevented us to make a proposal for the root distribution in the thermodynamic
limit. 

Inspite of that, the symmetry (\ref{SIM}) suggests that conclusions
for positive $\mathrm{U}$ may be achieved from the results obtained for
negative $\mathrm{U}$ without the need of novel calculations. To this
end we need however a correspondence connecting 
the {\bf low-lying} eigenstates for 
$\mathrm{U} >0$ with the {\bf low-lying} eigenstates for $\mathrm{U} <0$.
Inspecting the eigenvalues of the exact diagonalization 
we have been
able to uncover the following simple relation
between the lowest energies in the sector $n=1$,
\begin{equation}
\label{REL}
\mathrm{E}_1(\mathrm{U})
-\mathrm{E}_1(\mathrm{-U})=\frac{\mathrm{U}}{2} \mathrm{L}
\end{equation}
which was verified for even lattice sites up to $\mathrm{L}=12$ within
machine precision.

This result has prompted us to look for similar connection among
the ground state energies $\mathrm{E}_0(\mathrm{U})$ and
$\mathrm{E}_0(-\mathrm{U})$.
In Table (\ref{tab5}) we have exhibited our numerical results for finite-size estimates  
$\mathrm{F}_0(\mathrm{U})=\mathrm{E}_0(\mathrm{U})-\mathrm{E}_0(-\mathrm{U})-\frac{\mathrm{U}}{2}\mathrm{L}$
extracted from the exact diagonalization of the Hamiltonian.
Although $\mathrm{F}_0(\mathrm{U})$ is not exactly zero the deviation is very small 
indicating
that Eq.(\ref{REL}) should be valid for the ground state 
in the thermodynamic limit.
\begin{table}[ht]
\begin{center}
\begin{tabular}{|c||c|c|c|c|}
  \hline
$\mathrm{F}_0(\mathrm{U})$ & $\mathrm{U}=4$ & $\mathrm{U}=2\sqrt{3}$ & $\mathrm{U}=\sqrt{2}$ & $ \mathrm{U}=1$ \\ \hline \hline
4 & -0.0025~0261~7524 &  -0.0048~2183~0368  &  -0.0069~4028~6179  &  -0.0053~6392~2112   \\
6 & -0.0002~9857~8032 &  -0.0026~3677~4463  &  -0.0045~2586~3690  &  -0.0033~6949~6877  \\
8 &  0.0004~7434~8326 &  -0.0015~7527~8327  &  -0.0029~2025~0485  &  -0.0021~2556~6789  \\
10 & 0.0007~2533~0998 &  -0.0010~3343~6575   &  -0.0019~9639~6571  &  -0.0014~3687~8350  \\
12 & 0.0007~8126~9945 &  -0.0007~2631~8473  &  -0.0014~4003~0532  &  -0.0010~3046~6004      \\ \hline
\end{tabular}
\caption{ The finite-size sequences $\mathrm{F}_0(\mathrm{U})=\mathrm{E}_0(\mathrm{U})-\mathrm{E}_0(-\mathrm{U})-\frac{\mathrm{U}}{2}\mathrm{L}$.}
\label{tab5}
\end{center}
\end{table}

In any case these results tells us the lowest energy 
gap for both $\mathrm{U} <0$ and
$\mathrm{U} >0$ is certain of the same magnitude 
up to $\mathrm{L}=12$ sites. Assuming that this feature
remains for larger sizes and considering our 
findings for $\mathrm{U}>0$ we expect that   
the low-energy 
behaviour of the excitations 
of the spin chain
should be gapped for $|\mathrm{U}| > 2\sqrt{3}$ while being massless in the regime
$-2\sqrt{3} \leq \mathrm{U} \leq  2\sqrt{3}$. 

\section{Conclusions}
\label{sec4}
In this work we have studied the properties of the spectrum 
of a three-state vertex model based
on a $\mathrm{R}$-matrix which is non-additive with respect 
to the spectral variables.
Although the vertex weights lie on a curve with genus five 
a substantial part
of the structure of the transfer matrix eigenvalues was 
expressed by meromorphic
functions on an elliptic curve. This is because the eigenvalues depend on
very specific ratios of the weights representing a small subset of the
field of fractions of the genus five curve. We have noted that change on
the geometry of the weights has direct connection with 
the possibility of the presence
of different types of physical behaviour. This feature has been investigated 
in the case of an underlying spin one chain by means of analytical
and numerical methods. We find that the low-energy excitations 
can change from gapped to massless and the critical point separating
such phases is exactly the same found by geometrical considerations.

There are some open problems in the analysis of spin one chain which are
worth to pursue. The basis for an analytical study of the thermodynamic
limit properties in the massless regime is still lacking. This is important
for the derivation of the finite-size behaviour and
as consequence the underlying scaling dimensions. The observed correspondence
between low-lying states for positive and negative coupling deserves further
investigation. 
It is also of interest to establish functional relations
for the transfer matrix eigenvalues in the infinite system such
as the so-called inversion identities \cite{STRO}. This
method bypass the Bethe ansatz analysis providing us 
an alternative way 
to derive the free-energy and
the dispersion relation of
of the low-lying excitations \cite{KUM}. We hope that the results obtained 
here  will prompt further developments and interest in the study 
of spin chains originated out of non-additive $\mathrm{R}$-matrices.

\section*{Acknowledgments}
I thank T.S. Tavares and G.A.P. Ribeiro for fruitful discussions and help with some of the numerical part
of this work.
This work was supported in part  by the Brazilian Research Councils CNPq(2013/30329)

\addcontentsline{toc}{section}{Appendix A}
\section*{\bf Appendix A: Algebraic Bethe Ansatz }
%\label{APENA}
\setcounter{equation}{0}
\renewcommand{\theequation}{A.\arabic{equation}}

In what follows we summarized the structure of the eigenvectors 
of the transfer matrix (\ref{TRA},\ref{RMA},\ref{WEI}) within the general algebraic Bethe ansatz
formulation devised in the work \cite{CAMA}. In this framework the
eigenstates are expressed in terms of the elements of the 
monodromy matrix denoted here by ${\cal T}(\lambda)$. In the case of three state models it can
generically be written as,
\begin{equation}
{\cal T}(\lambda)=\left(\begin{array}{cccc}
                {\cal T}_{11}(\lambda) & {\cal T}_{12}(\lambda) & {\cal T}_{13}(\lambda) \\
                {\cal T}_{21}(\lambda) & {\cal T}_{22}(\lambda) & {\cal T}_{23}(\lambda) \\
                {\cal T}_{31}(\lambda) & {\cal T}_{32}(\lambda) & {\cal{T}}_{33}(\lambda) \\
                \end{array}\right),
\end{equation}
whose trace provides a representation for the transfer matrix, 
\begin{equation}
\mathrm{T}(\lambda)= 
{\cal T}_{11}(\lambda) + {\cal T}_{22}(\lambda) + {\cal T}_{33}(\lambda).
\end{equation}

A basic condition to diagonalize the transfer matrix by the 
algebraic Bethe ansatz
is the existence of a reference state $\ket{0}$ such that
${\cal T}(\lambda) \ket{0}$ gives a triangular matrix. In our case 
this pseudo 
vacuum  can
be taken as the tensor product of local ferromagnetic state, 
\begin{equation}
\ket{0}=\prod_{i=1}^{\mathrm{L}} \otimes \left(\begin{array}{c}
1 \\ 0 \\ 0 \end{array}\right)_{i}.
\end{equation}

Inspecting the action of the monodromy matrix on the 
pseudo vacuum $\ket{0}$ it is not 
difficult to derive the following identities,
\begin{equation}
{\cal T}_{11}(\lambda) \ket{0}= \left[a(\lambda,\mu)\right]^{\mathrm{L}},~~
{\cal T}_{22}(\lambda) \ket{0}= \left[{\bar{b}}(\lambda,\mu)\right]^{\mathrm{L}},~~
{\cal T}_{33}(\lambda) \ket{0}= \left[f(\lambda,\mu)\right]^{\mathrm{L}},~~
{\cal T}_{ij}(\lambda) \ket{0}= 0~~\mathrm{for}~~ i>j.
\end{equation}

The eigenstates of $T(\lambda)$ in the sectors with total 
magnetization $n=\mathrm{L}-m$ are 
constructed in terms of  a linear combination
of product of creation operators defined by 
the first row of the monodromy matrix. These states are viewed as   
$m$-particle states 
parameterized by the number of rapidities 
$\lambda_1,\dots,\lambda_m$ and they can be 
written as,
\EQ
\label{KET}
\ket{\Phi_m}=\phi_m(\lambda_1,\dots,\lambda_m) \ket{0}.
\EN

It turns out that the mathematical structure of the vector
$\phi_m(\lambda_1,\dots,\lambda_m) $ is given by means of a two-step 
recurrence relation, namely
\begin{eqnarray}
\label{eing}
\phi_m(\lambda_1,\dots,\lambda_m) &=&
{\cal T}_{12}(\lambda_1)
\phi_{m-1}(\lambda_2,\dots,\lambda_m) 
-{\cal T}_{13}(\lambda_1) \sum_{j=2}^{m} \varepsilon \frac{d(\lambda_1,\lambda_j)}{f(\lambda_1,\lambda_j)}
\phi_{m-2}(\lambda_2,\dots,\lambda_{j-1},\lambda_{j+1},\dots,\lambda_m) \nonumber \\
&\times& {\cal T}_{1,1}(\lambda_j) \prod_{\stackrel{k=2}{k \neq j }}^m \frac{a(\lambda_k,\lambda_j)}{\bar{b}(\lambda_k,\lambda_j)}
\theta_<(\lambda_k,\lambda_j), 
\end{eqnarray}
which is a non-trivial generalization of a previous work by Tarasov \cite{TARA}.

The permutation function 
$\theta_<(\lambda_{i},\lambda_{j})$ 
entering Eq.(\ref{eing}) is defined as,
\EQ
\theta_<(\lambda_{i},\lambda_{j})
= \begin{cases} \displaystyle
\theta(\lambda_{i},\lambda_{j}),~~~~\mbox{for} ~~ i < j \cr
\displaystyle 1, ~~~~\mbox{for} ~~ i \ge j.
\label{theta<}
\end{cases}
\EN
where the expression for $\theta(\lambda_{i},\lambda_{j})$ in terms of the weights has been given in Eq.(\ref{teta}).

We observe that the above vector can be related to each other under permutation  
of the rapidities,
\begin{equation}
\phi_m(\lambda_1,\dots,\lambda_j,\lambda_{j+1},\dots,\lambda_m) =
\theta(\lambda_{j},\lambda_{j+1})
\phi_m(\lambda_1,\dots,\lambda_{j+1},\lambda_{j},\dots,\lambda_m)
\end{equation}

The condition that the off-shell multiparticle states (\ref{KET},\ref{eing}) are in fact eigenstates
of the transfer matrix imposes restrictions on the rapidities $\lambda_1,\dots,\lambda_m$. These are the
Bethe equations already presented in the main text, see Eq.(\ref{BET}).

{}

\end{document}